# Understanding Polarization Properties of InAs Quantum Dots by Atomistic Modeling of Growth Dynamics

Vittorianna Tasco[1], Muhammad Usman[2], Maria Teresa Todaro[1], Milena De Giorgi[1], and Adriana Passaseo[1]

[1] *National Nanotechnology Laboratory, Istituto Nanoscienze CNR, Via Arnesano, 73100 Lecce, Italy*
[2] *Tyndall National Institute, Lee Maltings, Dyke Parade, Cork, Ireland*

**Abstract.** A model for realistic InAs quantum dot composition profile is proposed and analyzed, consisting of a double region scheme with an In-rich internal core and an In-poor external shell, in order to mimic the atomic scale phenomena such as In-Ga intermixing and In segregation during the growth and overgrowth with GaAs. The parameters of the proposed model are derived by reproducing the experimentally measured polarization data. Further understanding is developed by analyzing the strain fields which suggests that the two-composition model indeed results in lower strain energies than the commonly applied uniform composition model.

## INTRODUCTION

Understanding how the polarization response of semiconductor quantum dots (QDs) is related to their structural symmetry and composition profile is of critical importance for several applications ranging from optical communications to quantum information science. This objective is particularly challenging for III-V QDs, epitaxially formed by the self-assembling Stranski-Krastanov process, where atomic scale phenomena such as In-Ga intermixing and In-segregation effects during the capping and post-growth annealing processes [1-2] must be taken into account. Because of these processes, the actual InAs QD composition profile is far from being uniform throughout the nanostructure, as described by several theoretical studies and experimental papers [3]. Previous theoretical [4,5,6] studies of QD polarization response based on pure InAs QDs actually fail to reproduce the large values of the experimentally measured degree of polarization by using both k•p [5, 6] and atomistic tight binding methods [4]. Therefore, a more complex, structural and compositional model must be defined to find an accurate agreement for the optical behavior between theory and experiments, especially concerning polarization response.

In this work we study a double region QD scheme and how its geometrical and chemical characteristics influence the biaxial and hydrostatic strain profile inside the dots, thus allowing to understand and to engineer their optical behavior, mainly as far as polarization response is concerned.

## METHODOLOGIES

Theoretical modeling was performed using atomistic simulator NEMO 3-D, based on the valence force field (VFF) method for strain calculations and the twenty band $sp^3d^5s^*$ tight binding model for the electronic structure including both linear and nonlinear piezoelectric potentials [5]. The samples used in this study to test the model consist of single layer InAs QDs (2.8 MLs) grown by Molecular Beam Epitaxy (MBE), covered with a 20 nm GaAs cap. Their morphological and structural characterization by atomic force and transmission electron microscopy, shows a density around $4 \times 10^{10}$ dot/cm$^2$ and a dome-like shape with base diameter of 15 nm and height of 5 nm. The QD polarization ratio ($\rho$ = TM/TE) at room temperature was obtained by exciting the samples with a cw Ar+ laser ($\lambda$=514 nm) from the top and collecting photoluminescence (PL) signal from the cleaved edge after filtering with a linear polarizer.

## RESULTS AND DISCUSSION

The introduction of a two-region model is justified since pure InAs QD model fails to reproduce the experimental spectra. The TM/TE ratio is calculated to be 0.097 in this case, significantly lower than the measured value of 0.26. Moreover, the calculated PL spectra were also red-shifted by 123 nm. Even the assumption of an alloyed InGaAs model with uniform composition doesn't allow to reproduce both transition wavelength and polarization ratio [7].

In depicting our double region QD composition model we considered the microscopic influence of In–Ga intermixing and In segregation effects during the QD overgrowth with GaAs cap which lead to an In-rich region along the vertical direction and a decreasing In composition in the in-plane direction from the center to the border of the QD. The composition and geometry of the proposed model is shown by a schematic diagram in figure 1: the model is based on a high In-content inner core surrounded by a low In content outer shell.

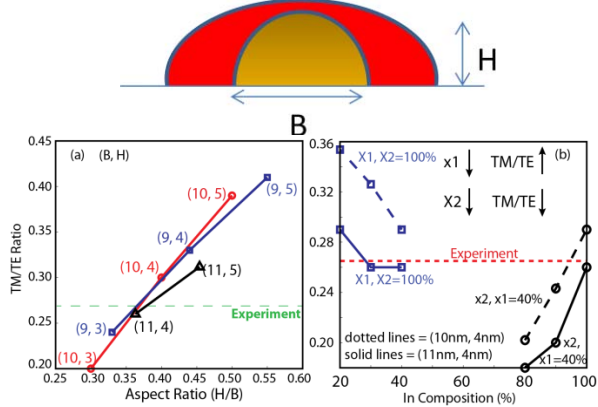

FIGURE 1. Top of the figure: double region quantum dot scheme: outer region in red with composition x1 and inner region in yellow with composition x2, diameter B and height H. Effects on the polarization ratio (TM/TE) of inner region aspect ratio (H/B) (a) and external and internal compositions (b) for two sets of (H, B).

The parameters defining our model are the external and internal compositions (x1 and x2, respectively) and the geometrical features of the inner region (height H and diameter B). Figure 1 shows the calculated effect of x1, x2 and of the inner core aspect ratio (AR = H/B) on the polarization ratio of the nanostructures. The reference polarization ratio of 0.26, as experimentally measured in our samples, is also indicated in the graphs. An increase of H for a fixed value of B reduces the biaxial strain close to the center of the QD thus providing a linear increase of $\rho$ with the aspect ratio (figure 1-a). But figure 1-a also evidences the impact of inner core diameter, in agreement with the lower sensitivity to AR expected in the hole energy levels of small QDs [8]. The two compositions were also examined (figure1-b), and the configuration which better fits experimental data (B~11 nm, H~ 4–5 nm, x1~40% and x2~90–100%) reproduces the effect of the anisotropic In–Ga intermixing behavior with a larger in-plane inter-diffusion region and a negligible inter-diffusion at the top (0–1 nm) where segregation predominated.

As a further confirmation of the validity of such a model, strain energy and strain field calculations were performed. Figure 2 plots the relaxed QD strain energies as a function of the total number of In atoms for different sets of double region configurations after the VFF minimization is achieved. The two cases that reproduced the experimental value of the TM/TE ratio are highlighted by arrows. It is worth noting that the pure InAs QD composition leads to a higher value of strain energy ($5100\times10^{-9}$ N.nm, not shown in figure 2).

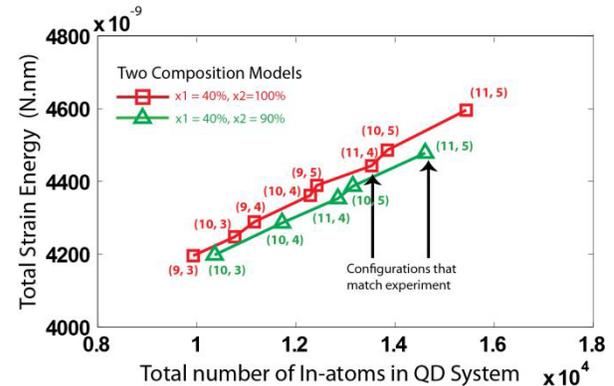

FIGURE 2. Total strain energy of the relaxed QD systems as a function of the total number of In atoms in the two-region system for different sets of parameters (B,H,x1,x2).

The calculation of the biaxial strain for the two composition model indicates that the biaxial strain component is reduced with respect to the pure InAs QD, thereby reducing the splitting between the heavy and light hole bands from 278 meV to 260 meV. This is consistent with the observed enhancement of the TM component.

This model can be used as a useful reference in designing QDs with specific optical properties: the growth dynamics can indeed be optimized in order to control atomic scale phenomena like intermixing and segregation, thus leading to properly engineered strain field profiles and consequently tuned optical behavior.